\documentstyle[prb,aps]{revtex}
\begin{document}

\title{Exact Optical Dielectric Function of a Nearly Homogeneous Solid}
\author{Girish S. Setlur and Yia-Chung Chang}
\address{Department of Physics and Materials Research Laboratory,\\
 University of Illinois at Urbana-Champaign , Urbana Il 61801}

\maketitle

\begin{abstract}
 Based on an exactly solvable model of a nearly homogeneous solid
 (jellium+phonons),
 we compute the optical dielectric function exactly at zero temperature
 in any number of spatial dimensions. It is argued as
 we have done elsewhere, that
 the answer for the optical dielectric function we obtain
 is in fact the exact answer for a real nearly homogeneous solid in the
 same limit in which RPA itself is exact.
\end{abstract}

\section{INTRODUCTION}

 In earlier preprints\cite{Setlur}, we introduced a scheme that involved
 replacing the electrons in a solid with sea-bosons.  This has the distinct
 advantage of rendering any problem that involves interacting electrons,
 exactly soluble in the RPA limit. In this article, we use the same approach
 to compute exactly the optical dielectric function of a nearly homogeneous
 solid.

\section{Model Hamiltonian}
The model hamiltonian of a nearly homogeneous solid may be written as,
\[
H = \sum_{ {\bf{k}}, {\bf{q}} }\omega_{ {\bf{k}} }({\bf{q}})
a^{\dagger}_{ {\bf{k}} }({\bf{q}})a_{ {\bf{k}} }({\bf{q}})
 + \sum_{ {\bf{q}} \neq {\bf{0}} }\frac{ v_{ {\bf{q}} } }{2V}
\sum_{ {\bf{k}}, {\bf{k}}^{'} }
[\Lambda_{ {\bf{k}} }({\bf{q}})a_{ {\bf{k}} }(-{\bf{q}})
 + \Lambda_{ {\bf{k}} }(-{\bf{q}})a^{\dagger}_{ {\bf{k}} }({\bf{q}}) ]
[\Lambda_{ {\bf{k}}^{'} }(-{\bf{q}})a_{ {\bf{k}}^{'} }({\bf{q}})
 + \Lambda_{ {\bf{k}}^{'} }({\bf{q}})a^{\dagger}_{ {\bf{k}}^{'} }(-{\bf{q}}) ]
\]
\begin{equation}
+ \sum_{ {\bf{q}} \neq {\bf{0}} }\frac{ M_{ {\bf{q}} } }{V^{\frac{1}{2}}}
\sum_{ {\bf{k}} }
[\Lambda_{ {\bf{k}} }({\bf{q}})a_{ {\bf{k}} }(-{\bf{q}})
 + \Lambda_{ {\bf{k}} }(-{\bf{q}})a^{\dagger}_{ {\bf{k}} }({\bf{q}}) ]
(b_{ {\bf{q}} } + b^{\dagger}_{ -{\bf{q}} })
+ \sum_{ {\bf{q}} }\Omega_{ {\bf{q}} }b^{\dagger}_{ {\bf{q}} }b_{ {\bf{q}} }
\end{equation}
Here as usual,
\begin{equation}
  \omega_{ {\bf{k}} }({\bf{q}}) = \Lambda_{ {\bf{k}} }(-{\bf{q}})
\frac{ {\bf{k.q}} }{m}
\end{equation}
and,
\begin{equation}
 \Lambda_{ {\bf{k}} }(-{\bf{q}}) = \sqrt{ n_{F}({\bf{k}}-{\bf{q}}/2)
(1 -  n_{F}({\bf{k}}+{\bf{q}}/2) ) }
\end{equation}
 and, $ n_{F}({\bf{k}}) = \theta(k_{f} - |{\bf{k}}|) $.
 This hamiltonian consists of the usual coulomb terms and the
 new ingredient here is the interaction with phonons. We assume that there is
 only one branch for simplicity, the optical phonon branch. 
 In order to diagonalise this, we proceed as usual, by introducing a Bogoliubov
 transformation \cite{Setlur}. Assume that the diagonalised form looks as
 follows,
\begin{equation}
H = \sum_{ {\bf{q}}, i}
\omega_{ {\bf{q}}, i }
d^{\dagger}_{ {\bf{q}}, i }d_{ {\bf{q}}, i }
\end{equation}
The relation between the dressed sea-displacements $ d_{ {\bf{q}}, i } $
 and the parent bosons may be deduced as follows,
The dressed sea-displacements and the original sea-displacements
 may be related by,
\begin{equation}
d_{ {\bf{q}}, i } = \sum_{ {\bf{k}} }
[d_{ {\bf{q}}, i }, a^{\dagger}_{ {\bf{k}} }({\bf{q}})]a_{ {\bf{k}} }({\bf{q}})
 - \sum_{ {\bf{k}} }
[d_{ {\bf{q}}, i }, a_{ {\bf{k}} }(-{\bf{q}})]a^{\dagger}_{ {\bf{k}} }
(-{\bf{q}})
+ [d_{ {\bf{q}}, i }, b^{\dagger}_{ {\bf{q}} }]b_{ {\bf{q}} }
 - [d_{ {\bf{q}}, i }, b_{ -{\bf{q}} }]b^{\dagger}_{ -{\bf{q}} }
\end{equation}
\begin{equation}
a_{ {\bf{k}} }({\bf{q}}) = \sum_{ i }
[a_{ {\bf{k}} }({\bf{q}}), d^{\dagger}_{ {\bf{q}}, i } ]d_{ {\bf{q}}, i }
 - \sum_{ i }
[a_{ {\bf{k}} }({\bf{q}}), d_{ -{\bf{q}}, i } ]
 d^{\dagger}_{ -{\bf{q}}, i }
\end{equation}
\begin{equation}
b_{ {\bf{q}} } = \sum_{ i }
[b_{ {\bf{q}} }, d^{\dagger}_{ {\bf{q}}, i } ]d_{ {\bf{q}}, i }
 - \sum_{ i }
[b_{ {\bf{q}} }, d_{ -{\bf{q}}, i }  ]
 d^{\dagger}_{ -{\bf{q}}, i }
\end{equation}
\begin{equation}
[d_{ {\bf{q}}, i }, a_{ {\bf{k}} }(-{\bf{q}})]
 = -(\frac{ \Lambda_{ {\bf{k}} }({\bf{q}}) }{
\omega_{ {\bf{q}}, i } + \omega_{ {\bf{k}} }(-{\bf{q}}) })
(\frac{ {\tilde{v}}_{ {\bf{q}}, i } }{V})
{\tilde{G}}_{2}({\bf{q}},i)
\end{equation}
\begin{equation}
[d_{ {\bf{q}}, i }, a^{\dagger}_{ {\bf{k}} }({\bf{q}})]
 = (\frac{ \Lambda_{ {\bf{k}} }(-{\bf{q}}) }{
\omega_{ {\bf{q}}, i } - \omega_{ {\bf{k}} }({\bf{q}}) })
(\frac{ {\tilde{v}}_{ {\bf{q}}, i } }{V})
{\tilde{G}}_{2}({\bf{q}},i)
\end{equation}
\begin{equation}
[d_{ {\bf{q}}, i }, b^{\dagger}_{ {\bf{q}} }]
 = (\frac{ M_{ {\bf{q}} } }{V^{\frac{1}{2}}})
(\frac{1}{ \omega_{ {\bf{q}}, i } - \Omega_{ {\bf{q}} } })
{\tilde{G}}_{2}({\bf{q}},i)
\end{equation}
\begin{equation}
[d_{ {\bf{q}}, i }, b_{ -{\bf{q}} }]
 = -(\frac{ M_{ {\bf{q}} } }{V^{\frac{1}{2}}})
(\frac{1}{ \omega_{ {\bf{q}}, i } + \Omega_{ {\bf{q}} } })
{\tilde{G}}_{2}({\bf{q}},i)
\end{equation}
here,
\begin{equation}
{\tilde{G}}_{2}({\bf{q}},i) = 
[(\frac{ M^{2}_{ {\bf{q}} } }{V})
\frac{ 4\omega_{ {\bf{q}}, i }\Omega_{ {\bf{q}} } }
{ (\omega^{2}_{ {\bf{q}}, i } - \Omega^{2}_{ {\bf{q}} })^{2} }
+ g^{-2}({\bf{q}},i)(\frac{ {\tilde{v}}_{ {\bf{q}}, i } }{V})^{2} ]
^{-\frac{1}{2}}
\end{equation}
\begin{equation}
{\tilde{v}}_{ {\bf{q}}, i } = 
v_{ {\bf{q}} } + M^{2}_{ {\bf{q}} }\frac{ 2\Omega_{ {\bf{q}} } }
{\omega^{2}_{ {\bf{q}}, i } - \Omega^{2}_{ {\bf{q}} } }
\end{equation}
\begin{equation}
 g^{-2}({\bf{q}},i) = \sum_{ {\bf{k}} }
[\frac{ n_{F}({\bf{k}}-{\bf{q}}/2 ) -  n_{F}({\bf{k}}+{\bf{q}}/2 ) }
{ (\omega_{ {\bf{q}}, i } -  \frac{ {\bf{k}}.{\bf{q}} }{m} )^{2} }]
\end{equation}
The eigenvalue $ \omega_{ {\bf{q}}, i } $ is given by,
\begin{equation}
\epsilon_{RPA}({\bf{q}}, \omega ) = 0
\label{EIGEN}
\end{equation}
and the RPA dielectric function with phonons is given by(analytic continuation
 is by replacing $ i\mbox{ }\omega \rightarrow \omega + i \mbox{ }0^{+}$). 
\begin{equation}
\epsilon_{RPA}({\bf{q}}, i\mbox{  }\omega) = 
1 + (\frac{ {\tilde{v}}_{ {\bf{q}}, \omega } }{V})
\sum_{ {\bf{k}} }
\frac{ n_{F}({\bf{k}} + {\bf{q}}/2) - n_{F}({\bf{k}} - {\bf{q}}/2) }
{ i\mbox{  }\omega - \frac{ {\bf{k}}.{\bf{q}} }{m} }
\end{equation}
\begin{equation}
{\tilde{v}}_{ {\bf{q}}, \omega } =
v_{ {\bf{q}} } - M^{2}_{ {\bf{q}} }\frac{ 2\Omega_{ {\bf{q}} } }
{\omega^{2} + \Omega^{2}_{ {\bf{q}} } }
\end{equation}
 The optical dielectric function is given by,
\begin{equation}
\epsilon(\omega) = 1 - \frac{ \omega^{2}_{p} }{ \omega^{2} }
 - \frac{ 4\pi }{ \omega^{2} }\pi(\omega)
\end{equation}
where $ \pi(\omega) $ is the retarded form of the current-current correlation
 function\cite{Mahan} :
\begin{equation}
 \pi(\omega) = - \frac{i}{V}\int_{-\infty}^{+\infty}
\mbox{ }dt\mbox{ }\theta(t-t^{'})\mbox{ }
exp(i\mbox{ }\omega\mbox{ }(t-t^{'}))
\langle [{\bf{J}}(t),{\bf{J}}(t^{'})] \rangle
\end{equation}
The total current is given by,
\begin{equation}
{\bf{J}} = e\sum_{ {\bf{k}}, {\bf{q}} }\Lambda_{ {\bf{k}} }(-{\bf{q}})
(\frac{ {\bf{q}} }{m})
\mbox{ }a^{\dagger}_{ {\bf{k}} }({\bf{q}})a_{ {\bf{k}} }({\bf{q}})
\end{equation}
If the current operator $ {\bf{J}} $ is independent of time then 
 $ \pi(\omega) = 0 $, and the imaginary part of the optical 
 dielectric function is identically zero, indicating that there
 is no absorption of radiation. This is going to happen
 in the case when phonons are absent and then we know that the total current
 is conserved in a homogeneous electron-gas. This fact is obvious in the 
 case when we use the undiagonalised form, but less so when we use the
 dressed-sea displacements. In order to verify this latter fact(namely that
 $ {\bf{J}} $ is time independent in the case when phonons are absent),
 let us write down the relevant quantity(possibly time dependent part
 of the total current) and argue that it is zero,
\begin{equation}
Part\mbox{ }of\mbox{ }{\bf{J}}
 = -e\mbox{ }\sum_{ {\bf{q}} }(\frac{ {\bf{q}} }{m})\sum_{ {\bf{k}} }
\sum_{i,j}[a_{ {\bf{k}} }({\bf{q}}), d^{\dagger}_{ {\bf{q}}, i }]
 [a_{ {\bf{k}} }({\bf{q}}), d_{ -{\bf{q}}, j }]
d^{\dagger}_{ {\bf{q}}, i }d^{\dagger}_{ -{\bf{q}}, j }
exp(i\mbox{ }(\omega_{ {\bf{q}}, i } + \omega_{ {\bf{q}}, j }) t )
\end{equation}
This is going to be zero if there is only one root to the eigenvalue equation, 
Eq.(~\ref{EIGEN}). This is the case when there are no phonons present.
 Even in more than one dimension, without phonons, the equation has only
 one root. This is the alternative (somewhat roundabout) explanation as to
 why the homogeneous electron gas absorbs no radiation. But when there are 
 phonons present, the eigenvalue equation has precisely two roots
 (no matter what the dimension is). This is why $ {\bf{J}} $ is now time
 dependent and this leads to a finite imaginary part of $ \pi(\omega) $.
 In order to demonstrate that there are precisely two roots of the equation
 Eq.(~\ref{EIGEN}), it is sufficient to solve the 1D case as this is likely
 to be indicative of the state of affairs in more than one dimension as well. 
 Unfortunately, the solution is not simple. One has to solve recursively the
 equation,
\begin{equation}
\omega = i\mbox{ }(\frac{|q|}{m})
\sqrt{ \frac{ (k_{f}+q/2)^{2} - (k_{f} - q/2)^{2}
exp(-(\frac{ 2\mbox{ }\pi\mbox{ }q }{m})
(\frac{1}{{\tilde{v}}_{q, \omega}})) }
{ 1- exp(-(\frac{ 2\mbox{ }\pi\mbox{ }q }{m})
(\frac{1}{{\tilde{v}}_{q, \omega}})) } }
\label{ROOT}
\end{equation}
The (postive) imaginary part of this root is the eigenvalue we are looking for.
The recursive solution obtained by replacing $ {\tilde{v}}_{q, \omega} $
 sucessively by first $ v_{q} $, and then later on by 
$ {\tilde{v}}_{q, \omega_{0}} $ where $  \omega_{0} $ is the zeroth order root
 and so on, is one such root. The other root is obtained by recursively solving 
 the same equation differently. Let us rewrite the recursion as,
\begin{equation}
\omega = i\mbox{ }
\sqrt{ \Omega^{2}_{ q }
 - \frac{ 2\mbox{ }M^{2}_{ q } \Omega_{ q } }
{ v_{q} - (\frac{2\pi\mbox{ }q}{m})
\{ ln(\mbox{ }S(q,\omega)\mbox{ }) \}^{-1} } }
\label{ROOT2}
\end{equation}
where,
\begin{equation}
S(q,\omega) = \frac{ 1 + (\frac{ k_{f}q }{m} - \epsilon_{q})^{2}/\omega^{2} }
{ 1 + (\frac{ k_{f}q }{m} + \epsilon_{q})^{2}/\omega^{2} }
\end{equation}
This may also be solved recursively, the zeroth order approximation being,
$ i\Omega_{q} $ and the rest as discussed before. 
 Thus we have eigenvalues, in the
 weak coupling regime they are close to $ k_{f}|q|/m $ and
 $ \Omega_{q} $ respectively. But in the strong coupling regime the answers are
 going to be different. The situation is likely to be similar in more than
 one dimension. Let us now compute the time evolved total current.
\begin{equation}
{\bf{J}}(t) = e \mbox{ }\sum_{ {\bf{k}}, {\bf{q}} }
(\frac{ {\bf{q}} }{m})\Lambda_{ {\bf{k}} }(-{\bf{q}}) 
a^{\dagger \mbox{ } t}_{ {\bf{k}} }({\bf{q}})a^{t}_{ {\bf{k}} }({\bf{q}})
\end{equation}
now,
\[
a^{t}_{ {\bf{k}} }({\bf{q}}) = 
\sum_{i = 1,2 }
[a_{ {\bf{k}} }({\bf{q}}), d^{\dagger}_{ {\bf{q}}, i }]d_{ {\bf{q}}, i }
e^{ -i\mbox{ }\omega_{ {\bf{q}}, 1 } t }
\]
\begin{equation}
 - \sum_{i = 1,2 }
[a_{ {\bf{k}} }({\bf{q}}), d_{ -{\bf{q}}, i }]
d^{\dagger}_{ -{\bf{q}}, i }e^{ i\mbox{ }\omega_{ {\bf{q}}, i } t }
\end{equation}
Therefore,
\begin{equation}
\langle [{\bf{J}}(t), {\bf{J}}(t^{'})] \rangle
 = \sum_{ {\bf{q}}, i \neq j }
F({\bf{q}}; i, j )
sin[(\omega_{ {\bf{q}}, i } + \omega_{ {\bf{q}}, j })(t^{'}-t)]
\end{equation}
\[
F({\bf{q}}; i, j )
 = (e^{2})(2\mbox{ }i)(\frac{ {\bf{q}}^{2} }{m^{2}})
\sum_{ {\bf{k}}, {\bf{k}}^{'} }\Lambda_{ {\bf{k}} }(-{\bf{q}})
\Lambda_{ {\bf{k}}^{'} }(-{\bf{q}})
[a_{ {\bf{k}} }({\bf{q}}), d_{ -{\bf{q}}, i }]
[a_{ {\bf{k}} }({\bf{q}}), d^{\dagger}_{ {\bf{q}}, j }]
[a_{ {\bf{k}}^{'} }({\bf{q}}), d^{\dagger}_{ {\bf{q}}, j }]
[a_{ {\bf{k}}^{'} }({\bf{q}}), d_{ -{\bf{q}}, i }]
\]
\begin{equation}
-(e^{2})(2\mbox{ }i)(\frac{ {\bf{q}}^{2} }{m^{2}})
\sum_{ {\bf{k}}, {\bf{k}}^{'} }\Lambda_{ {\bf{k}} }(-{\bf{q}})
\Lambda_{ {\bf{k}}^{'} }({\bf{q}})
[a_{ {\bf{k}} }({\bf{q}}), d_{ -{\bf{q}}, i }]
[a_{ {\bf{k}} }({\bf{q}}), d^{\dagger}_{ {\bf{q}}, j }]
[a_{ {\bf{k}}^{'} }(-{\bf{q}}), d^{\dagger}_{ -{\bf{q}}, i }]
[a_{ {\bf{k}}^{'} }(-{\bf{q}}), d_{ {\bf{q}}, j }]
\end{equation}
This leads us to the following formula for the retarded form of the
 current-current correlation function,
\begin{equation}
\pi(\omega) = (\frac{1}{V})\sum_{ {\bf{q}} }\sum_{ i \neq j }
(\frac{ F({\bf{q}}; i, j ) }{2\mbox{ }i})
\frac{1}{\omega - \omega_{ {\bf{q}}, i } -  \omega_{ {\bf{q}}, j } 
+ i\mbox{ }0^{+} }
\end{equation}
The formula that relates the refractive index and the absorption coefficient
 to the optical dielectric function has been given in the review by
 Haug and Schmitt-Rink \cite{Haug}.
\begin{equation}
\epsilon(\omega)^{\frac{1}{2}}
 = n(\omega) + i\frac{ c\mbox{ }\alpha(\omega) }{2\mbox{ }\omega}
\end{equation}
The imaginary and real parts are of $ \epsilon(\omega) $ are 
 easy to write down,
\begin{equation}
Re( \epsilon(\omega) ) = 1 - \frac{ \omega^{2}_{p} }{\omega^{2}}
 - \frac{ 4\pi }{ \omega^{2} }
(\frac{1}{V})\sum_{ {\bf{q}} }\sum_{ i \neq j }
(\frac{ F({\bf{q}}; i, j ) }{2\mbox{ }i})
{\mathcal{P}}
(\frac{1}{\omega - \omega_{ {\bf{q}}, i } -  \omega_{ {\bf{q}}, j } })
\end{equation}
\begin{equation}
Im( \epsilon(\omega) ) = \frac{ 4\pi^{2} }{ \omega^{2} }
(\frac{1}{V})\sum_{ {\bf{q}} }\sum_{ i \neq j }
(\frac{ F({\bf{q}}; i, j ) }{2\mbox{ }i})
\delta(\omega - \omega_{ {\bf{q}}, i } -  \omega_{ {\bf{q}}, j })
\end{equation}
The refractive index is given by,
\begin{equation}
n(\omega) = [(\frac{1}{2})[Re( \epsilon(\omega) ) 
 + ( (Re( \epsilon(\omega) ))^{2} + (Im( \epsilon(\omega) ))^{2} )
^{\frac{1}{2}} ] ]^{\frac{1}{2}} 
\end{equation}
\begin{equation}
\alpha(\omega) = \frac{ \omega \mbox{ }| Im(\epsilon(\omega)) | }
{ n(\omega) \mbox{ }c }
\end{equation}

\end{document}